\documentclass[12pt]{article}

\usepackage{amsmath}
\usepackage{graphicx,psfrag,epsf}
\usepackage{enumerate}
\usepackage{natbib}
\usepackage{amsmath}
\usepackage{amssymb}
\usepackage{algorithm}
\usepackage{bbm}
\usepackage{fancyhdr}
\usepackage{amsfonts}
\usepackage{subfig,xcolor}
\usepackage{hyperref}
\usepackage{amsthm}
\usepackage{wrapfig}
\usepackage{mathtools}  
\usepackage{enumitem}
\mathtoolsset{showonlyrefs}
\usepackage[noend]{algpseudocode}

\newcommand{\blind}{0}

\begin{document}

\def\spacingset#1{\renewcommand{\baselinestretch}%
{#1}\small\normalsize} \spacingset{1}


\if0\blind
{
  \title{\bf Cross-Validation Based Adaptive Sampling for Multi-Level Gaussian Process Models}
  \author{Louise M. Kimpton$^{\text{a}}$\thanks{
   The authors would like to thank EPSRC for their funding under the Excalibur grant number: EP/W007886/1}\hspace{.2cm}, James M. Salter$^{\text{a}}$, \\
  Tim J. Dodwell$^{\text{b}}$,  Hossein Mohammadi$^{\text{a}}$ \& Peter G. Challenor$^{\text{a}}$ \\
    $^{\text{a}}$Department of Mathematics and Statistics, University of Exeter \\
    $^{\text{b}}$digiLab and Institute for Data Science and AI, University of Exeter}
  \maketitle
} \fi

\if1\blind
{
  \bigskip
  \bigskip
  \bigskip
  \begin{center}
    {\LARGE\bf Cross-Validation Based Adaptive Sampling for Multi-Level Gaussian Process Models}
\end{center}
  \medskip
} \fi

\bigskip
\begin{abstract}
Complex computer codes or models can often be run in a hierarchy of different levels of complexity ranging from the very basic to the sophisticated. The top levels in this hierarchy are typically expensive to run, which limits the number of possible runs. To make use of runs over all levels, and crucially improve predictions at the top level, we use multi-level Gaussian process emulators (GPs). The accuracy of the GP greatly depends on the design of the training points. In this paper, we present a multi-level adaptive sampling algorithm to sequentially increase the set of design points to optimally improve the fit of the GP. The normalised expected leave-one-out cross-validation error is calculated at all unobserved locations, and a new design point is chosen using expected improvement combined with a repulsion function. This criterion is calculated for each model level weighted by an associated cost for the code at that level. Hence, at each iteration, our algorithm optimises for both the new point location and the model level. The algorithm is extended to batch selection as well as single point selection, where batches can be designed for single levels or optimally across all levels.
\end{abstract}

\noindent%
\vfill

\newpage
\spacingset{1} 

\section{Introduction}

Complex computer codes (or simulators) are key in describing many real-world applications where we are often interested in predicting physical outcomes \citep{Sacks1989}. These computer codes can often live in a hierarchy that can be run at different levels (or fidelities) of complexity ranging from the most basic to the very sophisticated \citep{Kattsov2013, Majda2010, Koziel2013}.  The levels of the model may be run at different resolutions depending on time or spatial resolution, or it may be possible to run these models at different levels of complexity where certain modelling components, assumptions or features can be removed or imposed, to generate a hierarchy of simpler models.  

The top levels of these computer codes typically have an extensive computational cost resulting in very few possible runs.  Although we are typically interested in gaining knowledge about these most complex levels, a larger number of cheaper runs of the code in the hierarchy still have important information that can help us to infer some underlying features of the full system we wish to model.

A common method for dealing with complex computer codes in a multi-fidelity framework was introduced by \cite{Kennedy2000}. In order to deal with expensive and complex computer models they use a surrogate model (or emulator) which is able to make predictions of model outputs whilst only requiring a small number of runs.  In this case, a Gaussian process emulator (GP) is used \citep{Haylock1996, Santner2003, OHagan2006}. Given $f(\cdot)$ represents the top level computer code that is used to describe the true observation, $y$, with error $\epsilon$, the equation for a two-level framework is:
\begin{equation}
    y = f(x) + \epsilon = \rho \cdot \tilde{f}(x) + \Big( \underbrace{f(x) - \rho \cdot \tilde{f}(x)}_{\delta(x)} \Big)  + \epsilon,
\end{equation}
where $ \tilde{f}(\cdot)$ gives the faster, lower level code, $\rho$ is a correlation parameter, and $\delta(\cdot)$ acts as a correction function to model the difference between the neighbouring levels. We reduce learning about $f$ to learning about two simpler functions ($\tilde{f}$ and $\delta$), which require fewer runs of the complex code whilst minimising  uncertainty.

The accuracy of the fit of any emulator (single or multi-level) is greatly dependent on the design of the training input points \citep{Sacks1989, Santner2003, Simpson2001, Currin1991}. It is important that the location of these points in the input space are space filling \citep{Joseph2016, Pronzato2012}, but also capture the important features of the physical system. This is more commonly known as the trade-off between exploration and exploitation, and is a topic that has been discussed extensively in current literature \citep{Liu2018, Garud2017}. Many authors use this as the basis of their design criteria. Examples of current widely used design methods include Latin hypercubes \citep{McKay1979}, minimum predictive uncertainty or mean squared error \citep{Sacks1989}, maximum entropy \citep{Shewry1987}, mutual information \citep{Krause2008} and expected improvement \citep{Schonlau1997} (from the Bayesian optimisation literature \citep{Jones2001, Brochu2010}), as well as other recommended references including \cite{Koehler2012} and \cite{Lam2008}, who give details on the pros and cons of each of these designs individually.  These methods of generating an optimum design fall into two main categories: one-shot designs or sequential designs \citep{Liu2018}. One-shot designs choose all samples at once and exhaust the computational budget immediately, whilst sequential designs select samples one (or more) at a time using information from the emulator or existing samples. 

The main difference between adaptive sampling for multi-level GPs as compared to single-level GPs is that not only must we select the optimum location of a new design point, we must also select the level of the model at which to run it. From a point of view of performing an adaptive design, the difference in levels is determined by their cost of running the code once. Typically this cost is defined as either the waiting time or the number of function evaluations in the computation. This is similar to the approach by \cite{Serani2019} and \cite{Pellegrini2016}, who base their selection on the emulator prediction uncertainty weighted by the ratio between neighbouring levels. Their approaches are limited by only considering a design based on either maximum uncertainty or expected improvement. This is similar to the kriging approaches from \cite{Huang2006}, \cite{Yi2021}, and \cite{LeGratiet2014} as well as an expected quantile approach from \cite{He2017}.  

This paper introduces a new method of adaptive sampling for multi-level Gaussian process emulators. The proposed method extends a single level version developed by \cite{Mohammadi2022}, based on calculating the normalised expected squared LOO (ES-LOO) error. This error can be defined at all unobserved locations and a new design point is chosen using expected improvement (EI) combined with a repulsion function. The repulsion function acts to push the selection away from already sampled points, to ensure the design is space filling. Further information about the ES-LOO algorithm is given in Section \ref{ASSL} and in \cite{Mohammadi2022}.

Due to the success of the single level ES-LOO algorithm, we believe the extension to multi-level applications will be vital in providing both a simple to use algorithm whilst maintaining an optimum balance between exploration and exploitation. In our method, we apply an adapted version of the ES-LOO adaptive sampling to each level of the model. We then weight each of the levels with an associated cost of running a design point at each of these model levels. We hence assign a cost $C^{(l)}$ to each of the hierarchy of levels $l=1, \ldots, L$ such that $C^{(1)} < C^{(2)} < \ldots < C^{(L)}$. Finally, both the level and design point are chosen where the weighted ES-LOO error is maximised. 

Other multi-level design approaches are fairly limited, where the majority focus on only a exploration criterion or assume a nested design across all levels of the hierarchy.  For example, \cite{Forrester2007} only perform an adaptive sampling design on the top level of the model, and then subsequently use this as the design for all levels.  \cite{Kuya2011} and \cite{Xiong2013} adapt this to perform a sequential design on the lowest level, and then take a subset from this for the design to be run at higher levels.  \cite{Stroh2022} aims to select new points based on maximising the ratio between the expected reduction of uncertainty and the cost of running the computer code.
Our criteria optimises for exploration and exploitation simultaneously, as well as offering a non-nested design approach. This allows us to develop a much simpler method where we can ignore any dependence structures and biases between the designs on separate layers. Our approach can then easily be applied to models with large numbers of hierarchical layers.

We believe the simplicity of our method is of great advantage when dealing with multi-level models. For example,  \cite{LeGratiet2015} introduce an approach to cokriging-based sequential design based on reducing the integrated mean squared error (IMSE). Calculating the IMSE is computationally expensive (especially in high dimensions) due to integrating the error across the input space. Hence, although the method captures uncertainty across the full space,  approximations to the integrals are often made. 

As an extension to the mutual information for computer experiments (MICE) algorithm \citep{Beck2016} noted in \cite{Mohammadi2022}, recent advances have been made to a multi-level framework \citep{Ehara2021}.  They specifically focus on a sampling method which allows for non-nested designs across levels whilst incorporating the total model error across all levels when optimising for the design selection. In order to take account of all error, they model all levels using a single GP, unlike the approach from \cite{Kennedy2000}. This can be beneficial, however the ES-LOO multilevel approach still maintains many of the advantages over MICE outlined in \cite{Mohammadi2022}, e.g., mutual information can be computational expensive in high dimensions as well as being limited since the criterion is based on a discrete representation. Although slower, we can optimise our criterion across the full input space.

The remainder of the paper is organised as follows. Section \ref{MLGP} gives an overview of multi-level Gaussian process emulators and Section \ref{ASSL} outlines the single level ES-LOO approach. Section \ref{ASML} extends this to adaptive design in a multi-level framework, including how to calculate each component and how to extend to batch design. Section \ref{Diss} adds important details about applying the methodology, and Section \ref{Exp1} gives an example.

\section{Preliminaries} \label{Bckg}

This section gives a brief overview of the background methodology required for our multi-level adaptive sampling method. We first outline multi-level Gaussian processes using the approach developed by \cite{Kennedy2000}, and then describe the adaptive sampling method for single level models introduced by \cite{Mohammadi2022}.

\subsection{Multi-level Gaussian Process Emulators} \label{MLGP}

Let the computer model be represented by the function $f : \mathcal{X} \mapsto \mathbb{R}$ with $\textbf{X} := (\textbf{x}_{1}, \ldots, \textbf{x}_{n})^{T} \in \mathcal{X} \subset \mathbb{R}^{p}$ being a set of $n$ inputs to the model in $p$ dimensions. The function $f(\cdot)$ hence maps the inputs $\textbf{X}$ to their associated outputs $\textbf{y} = (f(\textbf{x}_{1}), \ldots, f(\textbf{x}_{n}))^{T}$.  We can model $f$ as a Gaussian process $Z$ with training data $\{\textbf{X}, \textbf{y}\}$ such that:
\begin{equation}
Z(\textbf{x}) \sim GP(m(\textbf{x}), k(\textbf{x}, \textbf{x}')),
\end{equation}
where $m(\cdot)$ and $k(\cdot, \cdot)$ are predefined mean and covariance functions of the GP.  Using standard multivariate Normal distribution results, we can predict the function $f$ at a new location $\textbf{x}$ with posterior mean and covariance given as: 
\begin{equation}
\begin{split}
m^{*}(\textbf{x}) &= m(\textbf{x}) + k(\textbf{x},\textbf{X}) \textbf{K}^{-1} (\textbf{y} - m(\textbf{x})), \\
k^{*}(\textbf{x}, \textbf{x}') &= k(\textbf{x}, \textbf{x}') - k(\textbf{x},\textbf{X}) \textbf{K}^{-1} k(\textbf{X},\textbf{x}),
\end{split}
\end{equation}
where $\textbf{K}$ is an $n \times n$ covariance matrix with entries $K_{ij} = k(\textbf{x}_{i}, \textbf{x}_{j})$.

Let $l  = 1, \ldots, L$ represent each of the $L$ levels in a multi-level computer model, with the $l^{th}$ level denoted by $f^{(l)}$,  such that $l=1$ represents the most coarse or cheapest level, and $l=L$ represents the full complexity or top level. For each level of the model, let $\textbf{X}^{(l)} := ( \textbf{x}_{1}^{(l)}, \ldots, \textbf{x}_{n_{l}}^{(l)})$ be a set of $n_{l}$ inputs and $\textbf{y}^{(l)} = (f^{(l)}(\textbf{x}_{1}^{(l)}), \ldots, f^{(l)}(\textbf{x}_{n_{l}}^{(l)}))$ be the vector of associated outputs. 

When the top level of the model, $f^{(L)}(\textbf{x})$, is the function of interest, we can adopt the \cite{Kennedy2000} autoregressive model framework where, given the Markov-like property:
\begin{equation}
\text{cov} \left[ f^{(l)}(\textbf{x}), f^{(l-1)}(\textbf{x}') | f^{(l-1)}(\textbf{x}) \right] = 0,
\end{equation}
the $l^{th}$ level of the model can be stated as:
\begin{equation} \label{Eq1}
f^{(l)}(\textbf{x}) = \rho^{(l-1)} \cdot f^{(l-1)}(\textbf{x}) + \underbrace{\left( f^{(l)}(\textbf{x}) - \rho^{(l-1)} \cdot f^{(l-1)}(\textbf{x}) \right)}_{:= \delta^{(l)}(\textbf{x})}.
\end{equation}
Here, $\delta^{(l)}(\textbf{x}) = f^{(l)}(\textbf{x}) - \rho^{(l-1)} \cdot f^{(l-1)}(\textbf{x})$, for $l=2, \ldots, L$, acts as a corrector to the $(l-1)^{\text{th}}$ level to model the difference between the two neighbouring levels, and $\rho^{(l-1)}$ acts as a correlation parameter controlling how much $f^{(l-1)}$ relates to $f^{(l)}$, such that if $\rho^{(l-1)} = 0$, then $f^{(l-1)}$ and $f^{(l)}$ are independent. The top level $L$ of the model can hence be written as follows:
\begin{equation} 
\begin{split}
f^{(L)}(\textbf{x}) &= \rho^{(L-1)} \cdot f^{(L-1)}(\textbf{x}) + \left( f^{(L)}(\textbf{x}) - \rho^{(L-1)} \cdot f^{(L-1)}(\textbf{x}) \right) \\
 &= \rho^{(L-1)} \cdot f^{(L-1)}(\textbf{x}) + \delta^{(L)}(\textbf{x}) \\
&= \rho^{(1)} \ldots \rho^{(L-1)} f^{(1)}(\textbf{x}) + \rho^{(2)} \ldots \rho^{(L-1)} \delta^{(2)}(\textbf{x}) + \rho^{(3)} \ldots \rho^{(L-1)} \delta^{(3)}(\textbf{x}) + \ldots \\
& \hspace{1cm} + \rho^{(L-1)} \delta^{(L-1)}(\textbf{x}) + \delta^{(L)}(\textbf{x}) \\
&= P_{1} \cdot f^{(1)}(\textbf{x}) + \sum_{l=2}^{L} P_{l} \cdot \delta^{(l)}(\textbf{x}),
\end{split}
\end{equation}
where $P_{1} = \prod_{i=1}^{L-1} \rho^{(i)}$, $P_{l} =  \prod_{k=l}^{L-1} \rho^{(k)}$ for $l=2, \ldots, L-1$, and $P_L = 1$. 

Using this formulation, we can fit a multi-level Gaussian process to the top level of the model as follows:
\begin{equation} \label{MLGP1}
Z^{(L)}(\textbf{x}) = P_{1} \cdot Z^{(1)}(\textbf{x}) + \sum_{l=2}^{L} P_{l} \cdot Z_{\delta}^{(l)}(\textbf{x}).
\end{equation}
$Z^{(1)}(\textbf{x})$ is a GP for the cheapest level of the model with training data $D^{(1)} = \{\textbf{X}^{(1)}, \textbf{y}^{(1)}\}$, mean function $m^{(1)}(\cdot)$ and covariance function $k^{(1)}(\cdot, \cdot)$. $Z^{(l)}_{\delta}$ is a GP modelling $\delta^{(l)}$ as an independent Gaussian process with independent hyperparameters to $Z^{(l-1)}_{\delta}$.  The associated training data is $D^{(l)}_{\delta} = \{\textbf{X}^{(l)}, \textbf{y}^{(l)}_{\delta}\} \quad \mbox{with} \quad \textbf{y}_{\delta}^{(l)} = \big( f^{(l)}(\textbf{x}_{1}^{(l)}) - \rho^{(l-1)} \cdot f^{(l-1)}(\textbf{x}_{1}^{(l)}), \ldots, f^{(l)}(\textbf{x}_{n_{l}}^{(l)}) - \rho^{(l-1)} \cdot f^{(l-1)}(\textbf{x}_{n_{l}}^{(l)}) \big).$ The mean and covariance functions are given as $m^{(l)}_{\delta}(\cdot)$ and $k^{(l)}_{\delta}(\cdot, \cdot)$ respectively. Although $f^{(l-1)}(\textbf{x}^{(l)}_{i}) \in D_{\delta}^{(l)}$ for $i = 1,  \ldots, n_{l}$, we do not include $\textbf{x}_{i}^{(l)}$ and $f^{(l-1)}(\textbf{x}^{(l)}_{i})$ in the training data for the $(l-1)^{\text{th}}$ level, $D_{\delta}^{(l-1)}$. This means that we do not have nested designs for each level, such that $D_{\delta}^{(l)} \not\subset D_{\delta}^{(l-1)}$.  Although this does not make use of all available information at each level of the model, it does remove any potential bias that could arise from having an independent structure with non-independent nested data. This also adds flexibility in our sequential design approach, allowing a much cleaner and simpler result since no correlation between model levels needs to be considered. It is also noted that since each GP is independent, the choice of mean and covariance function can also differ by level. 

\subsection{Adaptive Sampling for Single Level Models} \label{ASSL}

Our new approach for multi-level adaptive sampling is based on a single-level version introduced by \cite{Mohammadi2022}, which combined expected squared leave-one-out error (ES-LOO) and pseudo expected improvement (PEI) to ensure a suitable trade-off between local exploitation and global exploration of the design.

Given a GP, $Z(\textbf{x})$, with training data $\{\textbf{X}, \textbf{y} \}$ with $\textbf{X} = (\textbf{x}_{1}, \ldots, \textbf{x}_{n})^{T}$ and $\textbf{y} = (f(\textbf{x}_{1}), \ldots, f(\textbf{x}_{n}))^{T}$, a leave-one-out cross-validation (LOO-CV) step is first carried out. Each input $\textbf{x}_{1}, \ldots, \textbf{x}_{n}$ is left out in turn, and a new GP, $Z_{-i}(\textbf{x})$, is estimated using this reduced training set and used to predict the output at the left out point. The predictive mean and variance of $Z_{-i}(\textbf{x})$ are denoted $m_{-i}(\textbf{x})$ and $s^{2}_{-i}(\textbf{x})$ respectively.

The expected squared LOO error (ES-LOO) is calculated to ensure that as well as accounting for the difference between the predictive mean and $f(\textbf{x}_{i})$ at $\textbf{x}_{i}$, the sensitivity of the emulator to the design points is accounted for, and is defined as:
\begin{equation}
\mathcal{E}(\textbf{x}_{i}) = \frac{\mathbb{E} \Big[ (Z_{-i}(\textbf{x}_{i}) - f(\textbf{x}_{i}))^{2} \Big]}{\sqrt{\text{Var} \Big[ (Z_{-i}(\textbf{x}_{i}) - f(\textbf{x}_{i}))^{2} \Big] }},
\end{equation}
where
\begin{align*}
\mathbb{E} \Big[ (Z_{-i}(\textbf{x}_{i}) - f(\textbf{x}_{i}))^{2} \Big] &= s^{2}_{-i}(\textbf{x}_{i}) + (m_{-i}(\textbf{x}_{i}) - f(\textbf{x}_{i}))^{2}, \\
\text{Var} \Big[ (Z_{-i}(\textbf{x}_{i}) - f(\textbf{x}_{i}))^{2} \Big] &= 2s^{4}_{-i}(\textbf{x}_{i}) + 4s^{2}_{-i}(\textbf{x}_{i})(m_{-i}(\textbf{x}_{i}) - f(\textbf{x}_{i}))^{2},
\end{align*}
as shown in \cite{Mohammadi2022}. The ES-LOO is defined over the full input space by modelling it as a Gaussian process with training data $\{\textbf{X}, \textbf{y}_{\mathcal{E}} \}$ with $\textbf{y}_{\mathcal{E}} = (\mathcal{E}(\textbf{x}_{1}), \ldots, \mathcal{E}(\textbf{x}_n))^{T}$. This GP, denoted by $Z_{\mathcal{E}}(\textbf{x})$, has predictive mean and variance given by $m_{\mathcal{E}}(\textbf{x})$ and $s_{\mathcal{E}}(\textbf{x})$ respectively. 

Instead of finding the maximum of $m_{\mathcal{E}}(\textbf{x})$ to propose a new design point,  expected improvement (EI) (as seen in Bayesian optimisation literature \citep{Schonlau1997, Jones2001,  Brochu2010}) can be used to ensure a maximum trade-off between exploration and exploitation in the design criteria. This is calculated as:
\begin{equation} \label{EEI}
EI(\textbf{x}) =
\begin{cases}
\left( m_{\mathcal{E}}(\textbf{x}) - \text{max}(\textbf{y}_{\mathcal{E}}) \right) \Phi(u) + s_{\mathcal{E}}(\textbf{x}) \phi(u) &\text{if} \, s_{\mathcal{E}}(\textbf{x}) > 0 \\
0 &\text{if} \, s_{\mathcal{E}}(\textbf{x}) = 0
\end{cases},
\end{equation}
where $u = \frac{m_{\mathcal{E}}(\textbf{x}) - \text{max}(\textbf{y}_{\mathcal{E}})}{s_{\mathcal{E}}(\textbf{x})}$ and $\phi(\cdot)$ and $\Phi(\cdot)$ represent the PDF and CDF of the standard Normal distribution respectively. 

To further ensure that there is a sufficient level of exploration in the design method, EI is extended to pseudo expected improvement (PEI) by multiplying $EI(\textbf{x})$ by a repulsion function (RF) defined as:
\begin{equation} \label{ERF}
RF(\textbf{x};\textbf{X}) = \prod_{i=1}^{n} \left[ 1 - \text{Corr} \left( Z_{\mathcal{E}}(\textbf{x}), Z_{\mathcal{E}}(\textbf{x}_{i}) \right) \right],
\end{equation}
where $\text{Corr}(\cdot,\cdot)$ is the correlation function of $Z_{\mathcal{E}}(\cdot)$. As it is typically not advantageous in most applications to place new design points in the furthest edges and corners of the input space, \cite{Mohammadi2022} also introduces a set of pseudo points $\textbf{X}_{p}$ into the repulsion function to reduce the chance of these types of points being selected.

The new design point $\textbf{x}_{n+1}$ is chosen with the following maximisation of the PEI:
\begin{equation} \label{SLPEI}
\textbf{x}_{n+1} = \underset{\textbf{x} \in \mathcal{X}}{\arg\max} \, \left[ PEI(\textbf{x}) := EI(\textbf{x}) \cdot RF(\textbf{x};\textbf{X} \cup \textbf{X}_{p}) \right].
\end{equation}
Further details, examples and an extension to batch design are included in \cite{Mohammadi2022}.

\section{Multi-Level Adaptive Sampling} \label{ASML}

The ES-LOO adaptive sampling approach selects the next best design point based on maximising the pseudo expected improvement across the input space. In this section, we propose methodology to extend this to adaptive sampling for a multi-level problem with $L$ levels, where the aim is to improve the fit of the multi-level Gaussian process prediction of the top, most expensive, level of the model.

As well as selecting the optimal location of the new point, we also need to consider at which of the $L$ levels of the hierarchy this should be run at. The associated cost for running $f^{(l)}$, the model at level $l$, is given as $C^{(l)}$ such that the hierarchy of models is defined by a hierarchy of simulation costs $C^{(1)} < \ldots < C^{(l)} < \ldots < C^{(L)}$.

Section \ref{sec:mlcrit} defines the proposed multi-level criterion. Sections \ref{sec:lev1} and \ref{sec:lev2} provide the calculations required for calculating the pseudo expected improvement for different levels of the hierarchy. Section \ref{sec:iter} shows the updates required after adding new points, and Section \ref{Batch} describes how to apply the criterion when batch designs are required.

\subsection{Proposed Multi-Level Criterion} \label{sec:mlcrit}

We propose that the new input point $\textbf{x}^{*}$ to be run at level $l^{*}$ satisfies the following criterion:
\begin{equation}  \label{EPEI}
(\textbf{x}^{*}, l^*) = \underset{\textbf{x} \in \mathcal{X}, \, l}{\arg\max} \, \left\{ \frac{1}{C^{(1)}} PEI^{(1)}(\textbf{x}) \cup \left( \bigcup_{l=2}^{L}  \frac{1}{C^{(l-1)} + C^{(l)}} PEI_{\delta}^{(l)}(\textbf{x}) \right) \right\},
\end{equation}
where $PEI^{(1)}(\cdot)$ is the pseudo expected improvement for level 1 and $PEI^{(l)}_{\delta}(\cdot)$ is the pseudo expected improvement for the $l^{\text{th}}$ difference Gaussian process, $Z_{\delta}^{(l)}$ (from Equation \eqref{MLGP1}). We seek to maximise the pseudo expected gain of each GP per unit cost, i.e. we find the input $\textbf{x}$ that maximises each $PEI^{(1)}$ and $PEI_{\delta}^{(l)}$ for $l = 2, \ldots, L$, and then select both the corresponding $\textbf{x}$ and level that gives the maximum $PEI$ across all levels, divided by the associated cost.

If $l^* = 1$, i.e. $\textbf{x}^{*}$ is selected to be run at level 1, then the training data for the GP at level 1 becomes:
$$
D^{(1)} = D^{(1)} \cup \{\textbf{x}^{*}, f^{(1)}(\textbf{x}^{*})\}.
$$
If $l^* > 1$, then we require two model runs in order to update the emulator fitted to the $l^{th}$ difference, with training data updated to be:
$$
D^{(l)}_{\delta} = D^{(l)}_{\delta} \cup \{\textbf{x}^{*}, f^{(l)}(\textbf{x}^{*}) - \rho^{(l-1)} \cdot f^{(l-1)}(\textbf{x}^{*}) \},
$$
hence giving the total cost of $C^{(l-1)} + C^{(l)}$ in Equation \eqref{EPEI} when $l > 1$.

Since we are now working in a multi-level framework, it is important in our design selection that we include any error from the GP across all levels of the model, and due to the form of the multi-level GP in \eqref{MLGP1}, the pseudo expected improvement for each level must be calculated differently to that in \cite{Mohammadi2022}. Hence, in place of the LOO error from the single-level ES-LOO design algorithm, we now have the total error on the $L^{\text{th}}$ level of the multi-level GP after performing a LOO-CV on each of the levels.

\subsection{Calculating PEI for level 1} \label{sec:lev1}

The approach works from the lowest level up, and hence we first find $PEI^{(1)}(\textbf{x})$. This is based on performing a leave-one-out cross-validation on the $1^{\text{st}}$ level Gaussian process only, $Z^{(1)}(\textbf{x}_{i})$, such that:
\begin{equation}
Z_{-i}^{(1)}(\textbf{x}_{i}) \sim \mathcal{N}(m_{-i}^{(1)}(\textbf{x}_{i}), (s_{-i}^{(1)})^{2}(\textbf{x}_{i})),
\end{equation}
where $Z^{(1)}_{-i}(\textbf{x}) = Z^{(1)}(\textbf{x}) | \textbf{y}^{(1)} \backslash \{ f^{(1)}(\textbf{x}_{i}) \}$, and where $m_{-i}^{(1)}$ and $(s_{-i}^{(1)})^{2}$ define the LOO predicted mean and variance respectively. From Equation \eqref{MLGP1}, the error on the $L^{\text{th}}$ level of the multi-level GP after performing this LOO on the $1^{\text{st}}$ level is hence given as:
\begin{equation}
e^{(L)}(\textbf{x}_{i}) = P_{1} \cdot  e^{(1)}_{LOO}(\textbf{x}_{i}) + \sum_{j=2}^{L} P_{j} \cdot e_{\delta}^{(j)}(\textbf{x}_{i}),
\end{equation}
with
\begin{equation}
e^{(1)}_{LOO}(\textbf{x}_{i}) = | m^{(1)}_{-i}(\textbf{x}_{i}) - f^{(1)}(\textbf{x}_{i})|,  \quad \mbox{and} \quad
e_{\delta}^{(j)}(\textbf{x}_{i}) = k^{(j)}_{\delta}(\textbf{x}_{i},\textbf{X}^{(j)}) (\textbf{K}_{\delta}^{(j)})^{-1} \textbf{y}_{\delta}^{(j)},
\end{equation}
where $\textbf{K}_{\delta}^{(j)}$ is the covariance matrix for $Z^{(j)}_{\delta}$ with entries $K^{(j)}_{\delta,ij} = k_{\delta}^{(j)}(\textbf{x}_{i}, \textbf{x}_{j})$. The error across all levels is additive since each $Z_{\delta}^{(j)}$ is independent and $D_{\delta}^{(j)} \not\subset D_{\delta}^{(j-1)}$.  Here, $e_{\delta}^{(j)}(\textbf{x}_{i})$ gives the error of $Z_{\delta}^{(j)}(\textbf{x}_{i})$ (for $Z_{\delta}^{(j)}(\textbf{x}) = Z^{(j)}(\textbf{x}) - \rho^{(j-1)} \cdot Z^{(j-1)}(\textbf{x})$). Since each error $e_{\delta}^{(j)}$ depends only on $\textbf{x}_{i}$ and is independent of the LOO-CV performed on the $1^{\text{st}}$ level of the model, we can define this to be equal to the full GP, $Z_{\delta}^{(j)}(\textbf{x}_{i})$. However,  since we are only interested in the variance of $Z_{\delta}^{(j)}(\textbf{x}_{i})$, we can simplify this GP to have zero mean so that:
\begin{equation}
Z_{e_{\delta}}^{(j)} (\textbf{x}_{i}) \sim \mathcal{N}(e_{\delta}^{(j)}(\textbf{x}_{i}), (s_{\delta}^{(j)})^{2}(\textbf{x}_{i})).
\end{equation}
Finally, due to standard Normal theory, we can therefore state the following GP for the total error on the top level of the model, $L$:
\begin{equation}
Z_{e}^{(L)}(\textbf{x}_{i}) = P_{1} \cdot \left(Z_{-i}^{(1)}(\textbf{x}_{i})-f^{(1)}(\textbf{x}_{i}) \right) + \sum_{j=2}^{L} P_{j} \cdot Z_{e_{\delta}}^{(j)}(\textbf{x}_{i}) 
\end{equation}
with distribution,
\begin{equation}
Z_{e}^{(L)}(\textbf{x}_{i})\sim \mathcal{N} \left( P_{1} \cdot \left(m_{-i}^{(1)}(\textbf{x}_{i})-f^{(1)}(\textbf{x}_{i}) \right) + \sum_{j=2}^{L} P_{j} \cdot e_{\delta}^{(j)}(\textbf{x}_{i}),  P_{1}^{2} \cdot  (s^{(1)}_{-i})^{2}(\textbf{x}_{i}) + \sum_{j=2}^{L} P_{j}^{2} \cdot  (s^{(j)}_{\delta})^{2}(\textbf{x}_{i}) \right).
\end{equation}

\cite{Mohammadi2022} shows that the LOO error does not always give the best results due to it only accounting for the difference between the true value at $\textbf{x}_{i}$ and the predicted mean when the input point is left out of the initial design.  Instead, the expected LOO is calculated which reduces these problems by also taking account of the prediction uncertainty. To find the normalised ES-LOO error $\mathcal{E}^{(1)}$ at the design point $\textbf{x}_{i}$ we have:
\begin{equation} \label{ENEI}
\mathcal{E}^{(1)}(\textbf{x}_{i}) = \frac{\mathbb{E}\left[ \left( Z_{e}^{(L)}(\textbf{x}_{i}) \right)^{2} \right]}{\sqrt{\text{Var} \left[ \left( Z_{e}^{(L)}(\textbf{x}_{i}) \right)^{2} \right]}},
\end{equation}
where:
\begin{equation}
\mathbb{E}\left[ \left(  Z_{e}^{(L)}(\textbf{x}_{i}) \right)^{2} \right] = P_{1}^{2} \cdot (s^{(1)}_{-i})^{2} (\textbf{x}_{i}) + \sum_{j=2}^{L} P_{j}^{2} \cdot (s^{(j)}_{\delta})^{2} (\textbf{x}_{i}) + \left( P_{1} \cdot \left( m_{-i}^{(1)}(\textbf{x}_{i}) - f^{(1)}(\textbf{x}_{i}) \right) + \sum_{j=2}^{L} P_{j} \cdot e_{\delta}^{(j)}(\textbf{x}_{i}) \right)^{2}
\end{equation}
and
\begin{align}
&\text{Var} \left[ \left( Z_{e}^{(L)}(\textbf{x}_{i}) \right)^{2} \right] = 2 \left( P_{1}^{2} \cdot (s^{(1)}_{-i})^{2} (\textbf{x}_{i}) + \sum_{j=2}^{L} P_{j}^{2} \cdot (s^{(j)}_{\delta})^{2} (\textbf{x}_{i}) \right)^{2} \\
& \hspace{1cm} + 4 \left( P_{1}^{2} \cdot (s^{(1)}_{-i})^{2} (\textbf{x}_{i}) + \sum_{j=2}^{L} P_{j}^{2} \cdot (s^{(j)}_{\delta})^{2} (\textbf{x}_{i}) \right) \times \left(P_{1} \cdot \left( m_{-i}^{(1)}(\textbf{x}_{i}) - f^{(1)}(\textbf{x}_{i}) \right) + \sum_{j=2}^{L} P_{j} \cdot e_{\delta}^{(j)}(\textbf{x}_{i}) \right).
\end{align}
Appendix A.1 shows how both of these expressions were calculated.

A GP $Z_{\mathcal{E}}^{(1)}$ is then fitted with training points $\{ \textbf{X}^{(1)}, \textbf{y}_{\mathcal{E}}^{(1)} \}$ and $\textbf{y}^{(1)}_{\mathcal{E}} = (\mathcal{E}^{(1)}(\textbf{x}_{1}^{(1)}), \ldots, \mathcal{E}^{(1)}(\textbf{x}^{(1)}_{n_{1}}))$ to predict $\mathcal{E}^{(1)}(\textbf{x})$ across the whole design space.  This is identical to that seen in the single-level methodology and so the pseudo expected improvement, $PEI^{(1)}(\textbf{x}) = EI^{(1)}(\textbf{x}) \cdot RF^{(1)}(\textbf{x})$,  can be calculated from the expected improvement and repulsion function as from Equations \eqref{EEI} and \eqref{ERF} respectively.

\subsection{Calculating PEI for level $l > 1$} \label{sec:lev2}

Let's now consider $PEI_{\delta}^{(l)}(\textbf{x})$ for $l = 2, \ldots, L$. The error on the $L^{\text{th}}$ level is now dependent on a LOO-CV performed on $Z_{\delta}^{(l)}$, where we have:
\begin{equation}
Z_{\delta,-i}^{(l)}(\textbf{x}_{i}) \sim \mathcal{N}(m_{\delta,-i}^{(l)}(\textbf{x}_{i}), (s_{\delta,-i}^{(l)})^{2}(\textbf{x}_{i}))
\end{equation}
for $Z_{\delta,-i}^{(l)}(\textbf{x}) = Z_{\delta}^{(l)}(\textbf{x}) | \textbf{y}_{\delta}^{(l)} \backslash \{ \delta^{(l)}(\textbf{x}_{i}) \}$ with $m_{\delta,-i}^{(l)}$ and $(s_{\delta,-i}^{(l)})^{2}$ defining the LOO predicted mean and variance respectively. The total error in Equation \eqref{MLGP1} with respect to this LOO on $Z_{\delta}^{(l)}$ now becomes:
\begin{equation}
e^{(L)}(\textbf{x}_{i}) = P_{1} \cdot  e^{(1)}(\textbf{x}_{i}) + P_{l} \cdot e_{LOO_{\delta}}^{(l)}(\textbf{x}_{i})+  \sum_{j=2, j\neq l}^{L} P_{j} \cdot e_{\delta}^{(j)}(\textbf{x}_{i}),
\end{equation}
with
\begin{equation}
\begin{split}
e^{(1)}(\textbf{x}_{i}) &= k^{(1)}(\textbf{x}_{i},\textbf{X}^{(1)}) (\textbf{K}^{(1)})^{-1} \textbf{y}^{(1)},  \\
e^{(l)}_{LOO_{\delta}}(\textbf{x}_{i}) &= | m^{(l)}_{\delta,-i}(\textbf{x}_{i}) - \delta^{(l)}(\textbf{x}_{i}) |,  \\
e_{\delta}^{(j)}(\textbf{x}_{i}) &= k_{\delta}^{(j)}(\textbf{x}_{i},\textbf{X}^{(j)}) (\textbf{K}_{\delta}^{(j)})^{-1} \textbf{y}_{\delta}^{(j)},
\end{split}
\end{equation}
where $\textbf{K}^{(1)}$ is the covariance matrix for $Z^{(1)}$ with entries $K^{(1)}_{ij} = k^{(1)}(\textbf{x}_{i}, \textbf{x}_{j})$, and with $\textbf{K}_{\delta}^{(j)}$ as above. For the same reasoning as above,  both of the errors $e^{(1)}$ and $e_{\delta}^{(j)}$ are generated through simplifying both $Z^{(1)}$ and $Z_{\delta}^{(j)}$ to have zero mean structures respectively such that we now have:
\begin{equation}
Z_{e}^{(1)}(\textbf{x}_{i}) \sim \mathcal{N}(e^{(1)}(\textbf{x}_{i}), (s^{(1)})^{2}(\textbf{x}_{i})) , \quad \mbox{and} \quad
Z_{e_{\delta}}^{(j)} (\textbf{x}_{i}) \sim \mathcal{N}(e_{\delta}^{(j)}(\textbf{x}_{i}), (s_{\delta}^{(j)})^{2}(\textbf{x}_{i})).
\end{equation}
We can therefore state the following GP for the total error on the top level $L$ of the model:
\begin{equation}
Z_{e}^{(L)}(\textbf{x}_{i}) = P_{1} \cdot Z_{e}^{(1)}(\textbf{x}_{i}) + P_{l} \cdot \left( Z_{\delta,-i}^{(l)}(\textbf{x}_{i}) - \delta^{(l)}(\textbf{x}_{i}) \right) + \sum_{j=2, j\neq l}^{L} P_{j} \cdot Z_{e_{\delta}}^{(j)}(\textbf{x}_{i}),
\end{equation}
with distribution,
\begin{equation}
\begin{split}
Z_{e}^{(L)}(\textbf{x}_{i}) &\sim \mathcal{N} \Bigg( P_{1} \cdot e^{(1)}(\textbf{x}_{i}) + P_{l} \cdot \left( m_{\delta,-i}^{(l)}(\textbf{x}_{i}) - \delta^{(l)}(\textbf{x}_{i}) \right) + \sum_{j=2, j \neq l}^{L} P_{j} \cdot e_{\delta}^{(j)}(\textbf{x}_{i}) \, , \\
& \hspace{3cm} P_{1}^{2} \cdot  (s^{(1)})^{2}(\textbf{x}_{i}) + P_{l}^{2} \cdot  (s^{(l)}_{\delta,-i})^{2}(\textbf{x}_{i}) + \sum_{j=2, j \neq l}^{L} P_{j}^{2} \cdot  (s^{(j)}_{\delta})^{2}(\textbf{x}_{i}) \Bigg),
\end{split}
\end{equation}
and the normalised ES-LOO $\mathcal{E}^{(l)}(\textbf{x}_{i})$ can then be found as from Equation \eqref{ENEI}. A GP $Z_{\mathcal{E}}^{(l)}$ is then fitted with training points $\{ \textbf{X}^{(l)}, \textbf{y}_{\mathcal{E}}^{(l)} \}$ and $\textbf{y}^{(l)}_{\mathcal{E}} = (\mathcal{E}^{(l)}(\textbf{x}_{1}^{(l)}), \ldots, \mathcal{E}^{(l)}(\textbf{x}^{(l)}_{n_{l}}))$ to predict $\mathcal{E}^{(l)}(\textbf{x})$ across the whole design space.

Finally, we can calculate the pseudo expected improvement, $PEI_{\delta}^{(l)}(\textbf{x}) = EI_{\delta}^{(l)}(\textbf{x}) RF_{\delta}^{(l)}(\textbf{x})$ in the exactly same way as both $PEI^{(1)}$ and the single-level ES-LOO in equations \eqref{EEI} and \eqref{ERF} respectively.  See Appendix A.2 for expressions for the expectation and variance used in the calculation of $\mathcal{E}^{(l)}(\textbf{x}_{i})$ above.

\subsection{Applying iteratively} \label{sec:iter}

As with the single level version, the multi-level ES-LOO sampling method can be applied iteratively to the current design, with the full process given in Algorithm 1. We have an initial estimate for $Z^{(L)}$ using an initial design, and then run the full methodology as described above, with
the calculations in the previous two sections enabling us to calculate all terms in \eqref{EPEI}, and hence select $(\textbf{x}^{*}, l^*)$ and perform the corresponding new run(s) of the computer model. Given the new run(s), the multi-level GP is updated, and the algorithm is run again, with the repulsion function now including this new point.

This process continues until a stopping criteria is reached.  This is problem specific but typically involves the amount of resources available. In many applications, we continue to increase our design until we either run out of time or computational resources. Alternatively, we can continue the iterative process until a desired accuracy is reached.

\subsection{Extension to Batch Design} \label{Batch}

It is often more efficient to run the computer code or function $f^{(l)}$ (at level $l$) for a set of input parameters instead of a single point. This is particularly efficient when parallel computing is available and the cost of evaluating $q$ runs of the model is equivalent to just running the model once.  The ES-LOO algorithm can be extended to batch designs by adapting the repulsion function \citep{Mohammadi2022}, where a batch design is defined as selecting $q > 1$ locations for evaluation at each iteration. Hence following the same notation as Section \ref{ASSL}, we can update the repulsion function for a batch of $q$ runs to be:
\begin{equation}
RF(\textbf{x}; \textbf{X} \cup \textbf{x}_{n+1} \cup \ldots \cup \textbf{x}_{n+q-1}) = \prod_{i=1}^{n+q-1} \left[ 1 - \text{Corr} \left( Z_{\mathcal{E}}(\textbf{x}), Z_{\mathcal{E}}(\textbf{x}_{i}) \right) \right]
\end{equation}
This updates the PEI, where the $q^{\text{th}}$ batch point is chosen via:
\begin{equation}
\textbf{x}_{n+q} = \underset{\textbf{x} \in \mathcal{X}}{\arg\max} \, \left[ PEI(\textbf{x}) = EI(\textbf{x}) \cdot RF(\textbf{x}; \textbf{X} \cup \textbf{x}_{n+1} \cup \ldots \cup \textbf{x}_{n+q-1}) \right].
\end{equation}

The main difference when considering a batch design for a multi-level model is that the additional $q$ points need not be allocated to the same level. This mainly depends on whether the cost of running the model at one chosen level $q$ times is the same as the cost of running the model once. It also depends on the user's preference. By using the repulsion function, we have the choice of whether to run all $q$ points at the same level, choose the optimal levels for each $q$ points, or decide on a combination of the two. For example, we could first choose $q/2$ points to be run at level $l$. Then for the remaining $q/2$ points we could run the full multi-level ES-LOO with the PEI corresponding to level $l$ removed from the selection in \eqref{EPEI}, selecting new points from the other levels only.

We can also take this approach to adapting the repulsion function if we are able to run several runs of a cheaper level (say on a laptop) whilst waiting for a more expensive level to run on a server or larger scale computer. A combination of further runs can be made whilst limited to the waiting cost availability. With all batch examples, it is important to update the correct repulsion function to ensure no duplicate points are chosen.

First consider the case when the entire batch of $q$ points will be evaluated at the same level. Initially, the full multi-level ES-LOO method is run to find the first new input point $\textbf{x}_{n_{l^*}+1}$ with its corresponding level $l^*$ from Equation \ref{EPEI}. The following $q-1$ points will then also be evaluated at level $l^*$ (and $l^*-1$ if $l^*>1$) and selected using the single level batch ES-LOO as above in Equation \eqref{SLPEI}. If $l^*=1$ the $q^{\text{th}}$ point $\textbf{x}_{n_{1}+q}$ is chosen as:
\begin{equation}
\textbf{x}_{n_{1}+q} = \underset{\textbf{x} \in \mathcal{X}}{\arg\max} \, \left[ PEI^{(1)}(\textbf{x}) = EI^{(1)}(\textbf{x}) \cdot RF^{(1)}(\textbf{x}; \textbf{X}^{(1)} \cup \textbf{x}_{n_{1}+1} \cup \ldots \cup \textbf{x}_{n_{1}+q-1}) \right],
\end{equation}
and if $l^*>1$, then the $q^{\text{th}}$ point $\textbf{x}_{n_{l^*}+q}$ is chosen as:
\begin{equation}
\textbf{x}_{n_{l^*}+q} = \underset{\textbf{x} \in \mathcal{X}}{\arg\max} \, \left[ PEI_{\delta}^{(l^*)}(\textbf{x}) = EI_{\delta}^{(l^*)}(\textbf{x}) \cdot RF_{\delta}^{(l^*)}(\textbf{x}; \textbf{X}^{(l^*)} \cup \textbf{x}_{n_{l^*}+1} \cup \ldots \cup \textbf{x}_{n_{l^*}+q-1}) \right].
\end{equation}

If a batch design is required but not all points are to be run at the same level, then each run in the batch will have a separate cost. The multi-level ES-LOO method used will be identical to that in Algorithm 1 with again exceptions to the repulsion function. The full algorithm is run $q$ times where the repulsion function is updated with each iteration in the same way to Equation \eqref{ERF}.  If an input $\textbf{x}^{*}$ is selected to be run at level $l^*$, then for the next iteration of the batch design, $\textbf{x}^{*}$ is included in the repulsion function at level $l^*$ to ensure no further points can be selected at this location. This becomes $RF^{(l^*)}(\textbf{x}; \textbf{X}^{(l^*)} \cup \textbf{x}^{*})$.

\begin{algorithm} \label{Alg1}
\caption{Multi-level sequential design approach using LOO cross-validation} 
\begin{algorithmic}[1] 
\State Generate initial design for each level $\textbf{X}^{(l)} = \{\textbf{x}_{1}^{(l)}, \ldots, \textbf{x}_{n_{l}}^{(l)} \}$ for $l = 1, \ldots, L$
\State Evaluate $\textbf{y}^{(1)} = f^{(1)}(\textbf{X}^{(1)})$ and  $\textbf{y}_{\delta}^{(l)} = \big( f^{(l)}(\textbf{X}^{(l)}) - \rho^{(l-1)} \cdot f^{(l-1)}(\textbf{X}^{(l)})  \big)$ for $l=2, \ldots, L$
\State Fit multi-level GP $Z^{(L)}(\textbf{x}) = P_{1} \cdot Z^{(1)}(\textbf{x}) + \sum_{i=2}^{L} P_{i} \cdot Z_{\delta}^{(i)} (\textbf{x})$ with training data $\{\textbf{X}^{(1)}, \textbf{y}^{(1)} \}$ and $\{\textbf{X}^{(l)}, \textbf{y}_{\delta}^{(l)} \}$ for $l = 2, \ldots, L$
\While{not stop} 
\For{$l = 1, \ldots, L$}
\For{$i = 1, \ldots, n_{l}$}
\State Calculate $\mathcal{E}^{(l)}(\textbf{x}_{i})$
\EndFor
\State Set $\textbf{y}_{\mathcal{E}}^{(l)} = \big( \mathcal{E}^{(l)} \big( \textbf{x}^{(l)}_{1} \big) , \ldots, \mathcal{E}^{(l)} \big( \textbf{x}^{(l)}_{n_{l}} \big) \big)^{T}$
\State Fit GP $Z_{\mathcal{E}}^{(l)}$ to $\{ \textbf{X}^{(l)}, \textbf{y}_{\mathcal{E}}^{(l)} \}$
\State Generate pseudo points: $\textbf{X}_{p}^{(l)}$ 
\EndFor
\State $\textbf{x}^{*} \leftarrow \underset{\textbf{x} \in \mathcal{X}}{\arg\max} \, \left\{ \frac{1}{C^{(1)}} PEI^{(1)}(\textbf{x}) \cup \left( \bigcup_{l=2}^{L}  \frac{1}{C^{(l-1)} + C^{(l)}} PEI_{\delta}^{(l)}(\textbf{x}) \right) \right\}$
\State $l^{*} \leftarrow$ level $l$ which gives $\text{max}(PEI^{(1)}(\textbf{x}^{*}), PEI_{\delta}^{(l)}(\textbf{x}^{*}))$ for $l = 2, \ldots, L$
\State $\textbf{X}^{(l^{*})} = \textbf{X}^{(l^{*})} \cup \{\textbf{x}^{*} \}$
\If{$l^{*} = 1$}
\State Evaluate $y^{*} = f^{(1)}(\textbf{x}^{*})$
\State Set $\textbf{y}^{(1)} = \textbf{y}^{(1)} \cup \{y^{*} \}$
\State $n_{1} \leftarrow n_{1} + 1$
\Else
\State Evaluate $y^{*} = \big( f^{(l^{*})}(\textbf{x}^{*})  - \rho^{(l^{*}-1)} \cdot f^{(l^{*}-1)}(\textbf{x}^{*}) \big)$
\State Set $\textbf{y}_{\delta}^{(l^{*})} = \textbf{y}_{\delta}^{(l^{*})} \cup \{y^{*} \}$
\State $n_{l^{*}} \leftarrow n_{l^{*}} + 1$
\EndIf
\State Update $Z^{(L)}(\textbf{x})$ with $\{\textbf{x}^{*}, y^{*} \}$
\EndWhile
\end{algorithmic}
\end{algorithm}

\section{Discussion on Methodology} \label{Diss}

The full methodology of the multi-level adaptive sampling approach using ES-LOO is given in Algorithm 1. Line 1 in the algorithm gives instruction to generate an initial design for each level $\textbf{X}^{(l)} = \{\textbf{x}_{1}^{(l)}, \ldots, \textbf{x}_{n_{l}}^{(l)} \}$ for $l = 1, \ldots, L$. In order to do this, we suggest that the design for the lowest level of the model $l=1$ is generated with a maximin Latin hypercube (LHC) design \citep{McKay1979} (or other space-filling design).  For higher level designs, $l>1$, we now have information on areas of potential high variance. We therefore suggest choosing the design using the standard ES-LOO technique from \cite{Mohammadi2022} on the previous level of the model. Using this, we can sequentially choose the next $n_{l}$ best locations for level $l-1$ and select these to be the design for level $l$. Our main aim is to reduce the overall uncertainty when modelling the top level. The only prior knowledge we have on the upper levels is the design of the lower levels and so placing a point in a higher level where a lower level has maximum uncertainty will help reduce the overall error.  This choice of starting designs is particularly useful when the levels are highly correlated or the levels differ through resolution. Alternatively, if it is important to ensure a highly exploratory design, then a LHC or similar for each level may be sufficient.

Although the standard ES-LOO method is already designed to have a sufficient level of exploration, we find in the multi-level case, it can be useful to add in occasional random points in the upper levels to ensure the space is fully explored. We suggest adding in an extra point roughly every 10 iterations, depending on both the total budget available and the dimension size. The multi-level GP can sometimes be falsely confident in certain areas of the input space where it assumes that the lower and upper levels are highly correlated.  However, large fluctuations can occur in the upper levels that can be very difficult to observe without some additional points in these areas. 

The use of pseudo points ensures that the boundaries of the input space are not favoured in the optimisation step, as in the single level version \citep{Mohammadi2022}. This is a common issue in many sequential design methods, where points on the edges of the space are prioritised despite potentially offering little to no information about the model itself. Additional pseudo points $\textbf{X}_{p}$ are introduced to the repulsion function ($RF(\textbf{x};\textbf{X}\cup\textbf{X}_{p})$) both in the corners of the input space and at the closest point on each face of the (rectangular) bounding input region to the current design points. These points are not evaluated at any level of the model, but are designed to repel the method from choosing extreme points.

Existing multi-level sequential designs, which primarily focus on choosing new points to minimise the predictive uncertainty of the Gaussian process, tend to place more points on input boundaries in higher levels \citep{Serani2019, Pellegrini2016}. Due to the increased cost of running a point at a higher level, new points are more likely to be chosen when the predictive uncertainty is highest. Since this is typically around the edges of the input space for Gaussian processes, we find that either more points are chosen in these areas, or that more points are placed in the lower levels when it would actually be beneficial to learn more about the higher levels. Hence, since there is more opportunity for a boundary point to be chosen in a multi-level model, we suggest the option of including more pseudo points along the edges of the input space, but this is dependent on how informative the boundary areas are in the given application. See Figures 4 and 5 in \cite{Mohammadi2022} for a visualisation of the use of pseudo points. We also point the reader to Section 3.2 of \cite{Mohammadi2022} where a lower bound $\theta_{\mathcal{E}}$ on the correlation length parameters of $Z_{\mathcal{E}}$ are stated. For the same reasons, this bound is important for the multi-level case and so we used, $\theta_{\mathcal{E}} = \sqrt{-0.5/\text{log}(10^{-8})}$.

Not only can our method be useful at dealing with expensive hierarchical models, but it can also be used to gauge the benefits in including the very top levels.  Consider the case when the top two levels are almost equivalent. First, we would notice that the correlation parameter, $\rho^{(L-1)}$, would be close to 1. Then, whilst running the algorithm, we would observe very few points selected to be run at the highest level with little difference being made in reducing any of the overall model error. This then indicates that there may be little point in incorporating the top level at all, and the information obtained from the next level down is sufficient. This can be greatly cost efficient in applications involving high scale computing. A few runs of the top model will however always be required to establish that there is indeed a high level of correlation between the levels.

\section{Example} \label{Exp1}

In this section, we present a simple toy example with two levels. There are two inputs to the model, $(x_{1}, x_{2}) \in \mathcal{X} = [0,1]^2$, with the two levels of the function given by:
\begin{align*}
f^{(1)}(x_{1}, x_{2}) &= x_{2} + x_{1}^{2} + x_{2}^{2} - \sqrt{2}, \\
f^{(2)}(x_{1}, x_{2}) &= f^{(1)}(x_{1}, x_{2}) + \text{sin}(2\pi x_{1}) + \text{sin}(4\pi x_{1} x_{2}).
\end{align*}
The level costs are taken to be $C^{(1)}=1$ and $C^{(2)}=8$.

\begin{figure}[t]
\centering
\includegraphics[scale=0.4]{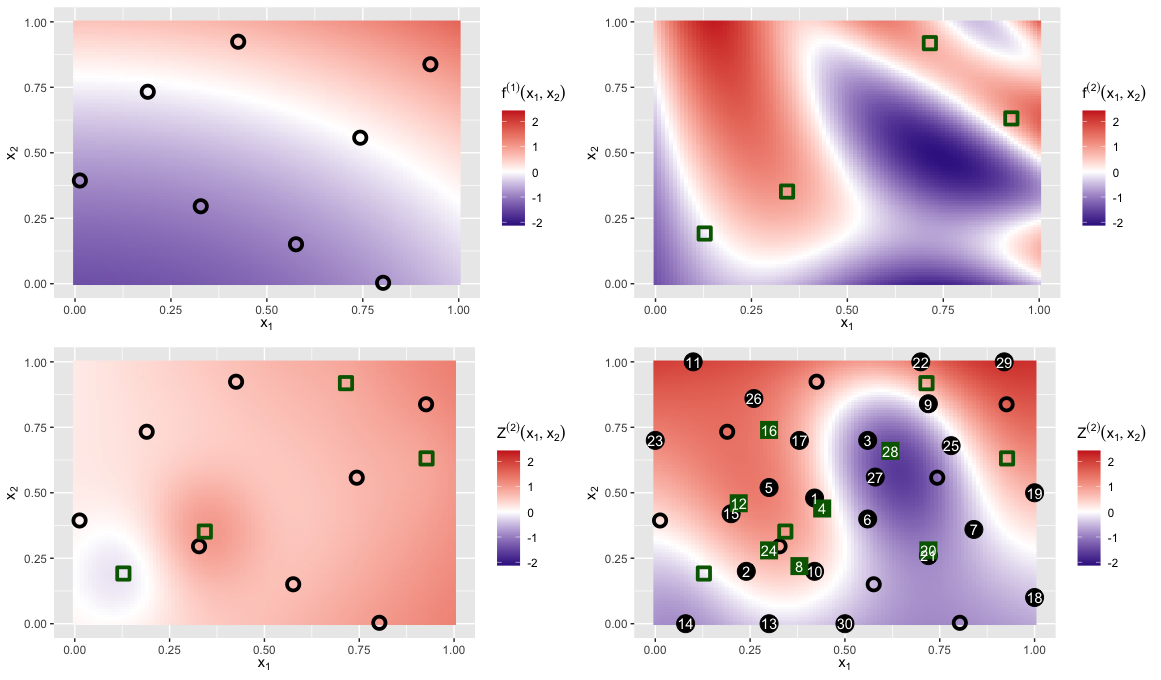}
\caption{Illustration of multi-level ES-LOO adaptive sampling approach on a two-dimensional example with two levels. Top left: true level 1 function, $f^{(1)}$, with initial design, $X^{(1)}$ (black circles). Top right: true level 2 (top level) function, $f^{(2)}$, with initial design $X^{(2)}$ (green squares). Bottom left: predicted multi-level output of $f^{(2)}$ using initial designs. Bottom right: Predicted multi-level output of $f^{(2)}$ after performing 30 iterations of the multi-level ES-LOO algorithm. New design points are also shown in the order that they were selected.}
\label{pic1a}
\end{figure}

We generate an initial design $\textbf{X}^{(1)}$ for the first level by selecting a Latin hypercube with 8 points. We then use the basic version of the ES-LOO algorithm with one level (from Section \ref{ASSL}) applied to level 1 to further select 4 points to become the initial design for level 2, $\textbf{X}^{(2)}$. The corresponding functions are then evaluated at these data points to generate the training data, $\{ \textbf{X}^{(1)}, \textbf{y}^{(1)} = f^{(1)}(\textbf{X}^{(1)}) \}$ and $\{ \textbf{X}^{(2)}, \textbf{y}^{(2)} = (f^{(2)}(\textbf{X}^{(2)}) - \rho^{(1)} \cdot f^{(1)}(\textbf{X}^{(2)}) \}$. The initial data and true functions ($f^{(1)}$ and $f^{(2)}$) are given in the top row in Figure \ref{pic1a}. Using this training data, we fit an initial multi-level GP:
\begin{equation}
Z^{(2)}(\textbf{x}) = \rho^{(1)} \cdot Z^{(1)}(\textbf{x}) + Z_{\delta}^{(2)}(\textbf{x}),
\end{equation}
with the correlation parameter set to $\rho^{(1)} = 1$. 

The predicted mean of $Z^{(2)}$ is given in the bottom left hand plot in Figure \ref{pic1a}, where the initial estimate is shown to be very poor. The accuracy of this prediction is determined by the normalised root mean squared error (NRMSE) \citep{LeGratiet2015}. Given a test set $D_{\tau} = \{ \textbf{X}_{\tau}, f^{(2)}(\textbf{X}_{\tau}) \}_{\tau=1}^{\tau=N}$, the NRMSE is given by:
\begin{equation}
\text{NRMSE} = \frac{\sqrt{\Big(\sum_{\tau=1}^{N}\big(m^{(2)}(\textbf{x}_{\tau}) - f^{(2)}(\textbf{x}_{\tau})\big)^{2}\Big)/N}}{\underset{\textbf{x}_{\tau} \in D_{\tau}}{\max} f^{(2)}(\textbf{x}_{\tau}) - \underset{\textbf{x}_{\tau} \in D_{\tau}}{\min} f^{(2)}(\textbf{x}_{\tau})},
\end{equation}
where $m^{(2)}(\textbf{x}_{\tau})$ is the expected mean output of the multi-level GP at input $\textbf{x}_{\tau}$ and we select $N=10000$ test points in a uniform grid across the input space. The NRMSE of the initial GP and corresponding design is 0.296.

We then apply the multi-level ES-LOO algorithm sequentially to give a total of 30 new design points. A total of 23 points were selected in level 1 and 7 points were selected in level 2, and these are shown in the bottom right hand plot in Figure \ref{pic1a} where they are labelled according to which order they were selected by the algorithm. Both the shape and colour represent which level of the model was selected to be run.  The predicted mean after including these new points into the GP training data is also shown in the bottom right of Figure \ref{pic1a}, where the NRMSE has dropped to 0.138 indicating an improved fit.   

\begin{figure}[t]
\centering
\includegraphics[scale = 0.45]{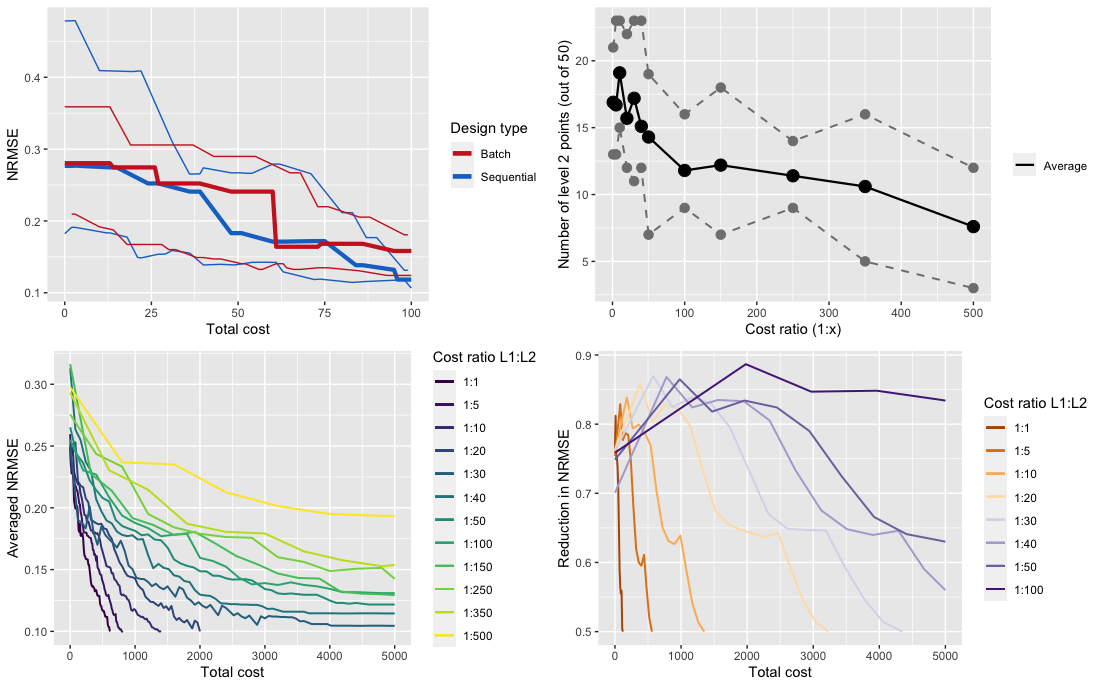}
\caption{For the example in Figure \ref{pic1a}, the multi-level ES-LOO algorithm is run for 10 different starting designs. Top left: multi-level ES-LOO algorithm is run for both the single sequential and batch versions and the total cost (up to 100) is plotted against the NRMSE. The NRMSE is given for the initial design in Figure \ref{pic1a} (thick lines), as well as 90\% confidence intervals across the 10 different starting designs (thin lines).
Top right: the cost ratio (of running level 1 against level 2) is set to range from 1:1 to 1:500, 50 iterations of the multi-level ES-LOO are performed, and the numbers of level 2 points selected are given.  Across the 10 designs, the average numbers of total level 2 points selected are shown in black along with 90\% confidence intervals (grey).
Bottom left: total running cost after selecting new design points against the average NRMSE across the 10 designs. Each line represents a different cost ratio (of running level 1 against level 2), ranging from 1:1 to 1:500.  
Bottom right: total running cost after selecting new design points against the reduction in NRMSE by modelling the top level with a multi-level GP as compared to a single-level GP. Each line represents a different cost ratio ranging from 1:1 to 1:100.}
\label{pic1b}
\end{figure}

In the top left plot in Figure \ref{pic1b} the NRMSE is plotted for both this single point sequential example and the batch design equivalent. For both algorithms, the ES-LOO algorithm is run until a budget of 100 is used, and we take the batch size to be 5. The thicker lines show the results for the initial design in Figure \ref{pic1a}, whilst the thinner lines show the 90\% confidence intervals for running the algorithm over 10 different starting designs. Both design types behave very similarly to each other, confirming that taking a batch approach to the multi-level ES-LOO method gives no disadvantages to the overall results.

To assess the accuracy and efficiency of the method further, we investigate the effects of the cost ratio between the two levels on both the proportion of level 2 points selected and the resulting NRMSE when this proportion differs.  It is expected that as the cost ratio of level 1 against level 2 reduces to 1:1, the number of level 2 points chosen will increase and that the NRMSE will also reduce at a faster rate. The example from Figure \ref{pic1a} is used with the exception of the initial starting design. To reduce the influence of the starting design, ten different initial designs are chosen where the method is applied to each. 

The top right plot in Figure \ref{pic1b} shows the effect that increasing the difference in cost between the two levels has on the number of level 2 points selected by the algorithm. For each of the ten starting designs, the multi-level ES-LOO algorithm is run for 50 iterations and the total number of level two points selected is recorded. As the cost of running the model at level 2 becomes much higher compared to level 1, the majority of points are run at the lower level. The cost of gaining that extra information is very high and is often not worth the increased cost. We see that when the ratio of costs between levels becomes close to 1:1, then around 15-20 level 2 points are selected. Although in this case we might assume that we'd be better off just running all points at level 2, we actually find that we would need fewer points to obtain a similar accuracy compared to just running the model at level 2. In this case, we still need many runs from level 1 to obtain a good estimate of the lower base level since the error in the GP is stacked across all levels. As the cost ratio exceeds 1:100, the number of level 2 points included remains fairly constant, indicating a minimum threshold of the number of top level points required to ensure a adequate fit of the multi-level GP.

The two bottom plots in Figure \ref{pic1b} illustrate the effects that adjusting the cost ratio has on the NRMSE scores.  The left plot shows the average NRMSE for various cost ratios ranging from 1:1 to 1:500. The multi-level ES-LOO algorithm is allowed to run so that along the x-axis, we have the total cost spent by the algorithm as we include new design points.  As the cost of level 2 decreases, it is clear that the NRMSE also decreases at a faster rate with respect to the total running cost. The right plot shows the reduction in NRMSE when applying the multi-level ES-LOO algorithm compared with just appling the single-level ES-LOO algorithm to the top level. For each cost ratio, as each algorithm is iterated, the total cost of the process is plotted against the difference in NRMSE between the multi-level and single level approaches. We can see that the difference in NRMSE is greater at the beginning, and then begins to level out as the total cost increases. This implies that the multi-level approach reaches an error threshold at a much faster rate, confirming that a lower cost is required for this algorithm to reach a certain level of accuracy. We can also see that the reduction in NRMSE is much higher when the difference in costs between levels is fairly low. Results are averaged over 10 starting designs as above.

\section{Conclusion} \label{conc}

We have presented our new approach to adaptive sampling for multi-level Gaussian processes. The algorithm is an extension to the single-level expected squared leave-one-out adaptive sampling approach introduced by \cite{Mohammadi2022}.  A new design point is chosen by maximising across the pseudo expected improvement calculated for each level weighted by the cost for running a point at that level in the computer code.  The cost is defined by the waiting time or the number of evaluations used for running the code.  For each level, the algorithm initially calculates a leave-one-out cross-validation on the initial design points to calculate the full error on the top level of the model. This includes a summation of both the LOO error on the selected level and the GP errors on all other levels. A second GP is then fitted to the normalised expected squared LOO error to estimate the error at unobserved locations.  We then calculate the pseudo expected improvement by multiplying the expected improvement by a repulsion function.  We have shown that the method can be used in a batch mode which can be adapted to suite the user's needs. 

Multi-level design is an area of great interest in current literature due to the increasing use of high scale computing for simulating physical systems. There have been previous attempts by other authors to solve this problem, but we believe there is not a current method that is quick and easy to use whilst maintaining an optimum balance between exploration and exploitation. Many other methods only focus on variance based criteria, or do not consider the full input space across all model layers in the design algorithm. We also have defined that the designs for each of the levels are not nested and that the algorithm accounts for all error included in the multi-level Gaussian process. This allows us to have a much computationally simpler method which ignores any dependence structures and biases between the designs on separate layers. As well as this, our method can easily be applied to high dimensions and has a criterion that can be fully optimised and is not limited to being based on a discrete representation.

There are a few areas where our method can be extended.  Firstly we have noticed a gap in the literature for adaptive sampling methods for stochastic models. Since the ES-LOO algorithm only depends on a few runs of the computer code, we believe that it should be able to be extended to cases where the variance varies across the input space. The main issues we envision are down to the replication involved in stochastic emulation. At each stage, we will have the choice between selecting a new design point, or selecting an existing point to replicate. Secondly, we could consider cases where the cost of each level is not consistent for all input points. This is quite common in practice, hence it would be interesting to see how the ES-LOO algorithm could be adapted to suit these situations.

\bibliographystyle{abbrvnat}
\bibliography{MLDbibF}
\end{document}